\documentclass{article}

\usepackage{PRIMEarxiv}

\usepackage[utf8]{inputenc} 
\usepackage[T1]{fontenc}    
\usepackage[hidelinks]{hyperref}       
\usepackage{url}            
\usepackage{booktabs}       
\usepackage{amsfonts}       
\usepackage{nicefrac}       
\usepackage{microtype}      
\usepackage{lipsum}
\usepackage{float}
\usepackage{multirow}
\usepackage{fancyhdr}       
\usepackage{graphicx}       
\graphicspath{{media/}}     
\usepackage{array}
\usepackage{makecell}
\title{A Scoping Review of AI-Driven Digital Interventions in Mental Health Care: Mapping Applications Across Screening, Support, Monitoring, Prevention, and Clinical Education
}

\author{
  Yang Ni \\
 School of International and Public Affairs \\
 Columbia University \\
 New York\\
  \texttt{yang.ni2@columbia.edu} \\
   \And
  Fanli Jia \\
  Department of Psychology \\
  Seton Hall University \\
  South Orange\\
  \texttt{fanli.jia@shu.edu} \\
}

\begin{document}
\maketitle

\begin{abstract}
Background/Objectives: Artificial intelligence (AI)-enabled digital interventions are increasingly used to expand access to mental health care. This PRISMA-ScR scoping review maps how AI technologies support mental health care across five phases: pre-treatment (screening), treatment (therapeutic support), post-treatment (monitoring), clinical education, and population-level prevention. Methods: We synthesized findings from 36 empirical studies published through January 2024 that implemented AI-driven digital tools, including large language models (LLMs), machine learning (ML) models, and conversational agents. Use cases include referral triage, remote patient monitoring, empathic communication enhancement, and AI-assisted psychotherapy delivered via chatbots and voice agents. Results: Across the 36 included studies, the most common AI modalities included chatbots, natural language processing tools, machine learning and deep learning models, and large language model-based agents. These technologies were predominantly used for support, monitoring, and self-management purposes rather than as standalone treatments. Reported benefits included reduced wait times, increased engagement, and improved symptom tracking. However, recurring challenges such as algorithmic bias, data privacy risks, and workflow integration barriers highlight the need for ethical design and human oversight. Conclusion: By introducing a four-pillar framework, this review offers a comprehensive overview of current applications and future directions in AI-augmented mental health care. It aims to guide researchers, clinicians, and policymakers in developing safe, effective, and equitable digital mental health interventions.
\end{abstract}


\section{Introduction}
In recent decades, mental health disorders have surged despite economic and technological progress, with barriers such as stigma, cost, and professional shortages continuing to hinder treatment access \cite{1,2,3}. Digital health technologies offer promise in addressing these challenges \cite{4}, with teletherapy, mental health apps, and computerized cognitive behavioral therapy showing effectiveness \cite{5}. The COVID-19 pandemic further accelerated the use of digital mental health tools, revealing unmet needs at the intersection of technology and psychotherapy \cite{6,7}. This growing demand has sparked increased interest in the role of conversational artificial intelligence in mental health applications \cite{8,9}, with research exploring its potential to enhance psychotherapy and improve care delivery \cite{10,11,12,13}. Researchers have increasingly focused on leveraging conversational artificial intelligence to support psychotherapy effectiveness, as the mental health sector grapples with rising demand and the need for innovative solutions \cite{12,13,14,15,16}. Since the public release of ChatGPT (Version GPT-4, United States of America) in early 2023, it has become the first conversational artificial intelligence tool to achieve global mainstream use, reshaping approaches to learning, communication, and problem solving \cite{8,17}. Research on ChatGPT has expanded rapidly across disciplines, especially in education, medicine, and psychology, highlighting its growing potential to advance mental health services \cite{10,11,12,13}.

Artificial intelligence-driven digital interventions in mental health refer to software systems or mobile applications that embed artificial intelligence techniques to deliver, support, or evaluate mental health services \cite{6,12}. These include conversational artificial intelligence agents that interact with users through natural language, ranging from simple FAQ style or rule-based chatbots to more advanced multi-turn dialogue systems capable of handling complex communication tasks \cite{18,19}. Natural language processing techniques enable these agents to parse user input, detect sentiment, and extract key emotional cues \cite{20}. Machine learning models, such as classification and regression algorithms, and deep learning networks, such as convolutional and recurrent neural networks, power predictive and monitoring tools to classify diagnoses, forecast risks, and tailor treatment recommendations based on user data \cite{20}. Large language models such as GPT and BERT, which belong to a subclass of deep learning models built on transformer architectures with self-attention mechanisms, expand capabilities by generating and comprehending coherent and context-rich text, opening new possibilities for nuanced therapeutic dialogue and personalized content creation \cite{18,19}. Artificial intelligence has emerged as a promising tool to augment human therapists \cite{12}, although its adoption challenges traditional care models and raises concerns regarding efficacy, ethics, privacy, and the interpretation of human mental health experiences \cite{21}. This review aims to provide a comprehensive analysis of conversational artificial intelligence in mental health care by mapping empirical evidence across different clinical phases. It offers insights into current applications, challenges, and future opportunities. Although previous reviews have focused narrowly on large language model capabilities, ethical concerns, or specific generative artificial intelligence applications, this scoping review provides a broader synthesis by mapping artificial intelligence-driven digital interventions across five clinical phases of mental health care: (1) pre-treatment, (2) treatment, (3) post-treatment, (4) clinical education, and (5) general improvement and prevention. We organize this landscape into a unified life cycle framework grounded in empirical evidence. Four research questions guide our analysis:

Research Question 1 (RQ1): Within each clinical phase, which artificial intelligence modalities (rule-based chatbots, natural language processing, machine learning or deep learning models, and large language models) power digital interventions, and what evidence exists regarding their efficacy and limitations?

Research Question 2 (RQ2): For each phase, which artificial intelligence-driven tools demonstrate the greatest impact, and what performance metrics and barriers have been reported?

Research Question 3 (RQ3): What are the general strengths, weaknesses, opportunities, and threats of utilizing artificial intelligence-driven interventions in mental health care both overall and within each clinical phase?

Research Question 4 (RQ4): Which artificial intelligence technologies and applications appear most mature today, which emerging trends warrant priority research, and which technical, clinical, and policy challenges must be addressed to advance artificial intelligence-driven digital interventions?

By linking conceptual insights with empirical outcomes, this review complements and extends prior work to provide an integrated reference for research, practice, and policy. Finally, it aims to equip future researchers and practitioners to identify promising development pathways and positions in this review as a key reference for understanding the full spectrum of artificial intelligence applications in mental health care.

\section{Methods}
\subsection{Research Aims}
Various types of AI technologies are utilized within the broad context of mental health care. Many of these technologies are specifically linked to the conversational AI interface, which engages users to provide a wide range of support. Various technologies, including AI chatbots, different language models, prediction modeling, sentiment analysis, and recommender systems, have been implemented into health care settings \cite{22,23,24,25}. These technologies are making significant advancements in mental health care by improving diagnostic accuracy, enhancing personalized treatment, providing insights and recommendations to clinicians, tailoring services to individual needs, and offering accessible and cost-effective mental health support to everyone \cite{26,27,28,29}. As the field of computer science continues to progress, there is a transformative opportunity for the mental health care field to understand and apply these technologies to their services effectively.

However, it is crucial to approach this integration thoughtfully, maintaining standards of care and prioritizing patient-centric approaches. There is a need for a scoping review examining how different AI technologies are being used in mental health, their impacts, ethical considerations, and practical aspects of combining AI with human care across various settings. This review aims to present a framework for the existing state of integrating AI into mental health services in a manner that maximizes benefits and minimizes risks. The summarization can serve as an overview for those interested in researching and developing solutions for these issues.

To realize their full promise, integration must be guided by evidence on efficacy, ethical safeguards, and alignment with clinical workflows. Accordingly, this scoping review maps the current landscape of AI-driven digital interventions in mental health, proposes a four-pillar mapping to organize empirical findings, and identifies practical barriers and enablers to maximize benefits and minimize risks. Our primary goals are to chart which AI modalities are deployed in each care phase, summarize their proven outcomes and reported limitations, and outline strategic directions for future research and policy.

\subsection{Design and Scope of the Study}
This scoping review adhered to the PRISMA-ScR guidelines \cite{30} to ensure a rigorous, transparent, and reproducible methodology for mapping the use of AI-driven digital interventions in mental health care. We define our scope as all conversational AI agents (from rule-based/FAQ chatbots to ML-powered multi-turn systems and transformer-based LLMs) and related predictive/monitoring models (NLP and ML/DL algorithms) deployed across five phases: (1) pre-treatment (screening and triage), (2) treatment (therapeutic support), (3) post-treatment (follow-up and monitoring), (4) clinical education, and (5) general improvement and prevention. Conducting this review, we examined how each technology is applied, its demonstrated outcomes (e.g., accuracy, engagement, and health gains), and its limitations, thereby offering a unified life cycle framework for AI in mental health.

\subsection{Identification and Selection of Studies}
Our search strategy encompassed empirical research reports and publications up to January 2024, focusing on the application and efficacy of artificial intelligence technologies in mental health care. The search included multiple databases using keywords and phrases related to conversational artificial intelligence, machine learning, and mental health, such as “conversational AI and mental health”, “ChatGPT psychotherapy”, “AI counseling”, “AI psychotherapy”, “AI counselor”, and “machine learning mental health”.

Initially, the comprehensive search yielded 1674 records. After removing 724 duplicates and excluding 804 articles that did not meet predefined criteria, 146 records remained for further evaluation. We retrieved 143 full-text reports for in-depth assessment.

Inclusion criteria targeted empirical research reports and publications published in English that examined the application of artificial intelligence technologies in mental health contexts. We excluded non-empirical works, such as literature reviews, editorials, and opinion pieces, as well as studies focusing on non-AI technologies or those outside the scope of mental health. Following screening, we excluded 82 non-empirical studies, 4 preprints that did not meet inclusion criteria, and 20 articles that fell outside our scope.

The first author independently conducted the search and data extraction using a custom-designed template to systematically capture key information from each study, including objectives, AI technologies used, main findings, and conclusions. Discrepancies were resolved through discussion to ensure consensus with the second author. Data were organized by clinical phase, AI modality, reported outcomes, and identified limitations. In total, 36 studies met the inclusion criteria and formed the core sample for this scoping review (Figure 1). These studies represent the most relevant and methodologically sound research currently available on conversational AI in mental health care. Owing to the heterogeneity of study designs and outcomes, we did not conduct a risk-of-bias appraisal or quantitative synthesis.

\newpage
\begin{figure}[H]
    \centering
    \includegraphics[width=0.6\textwidth]{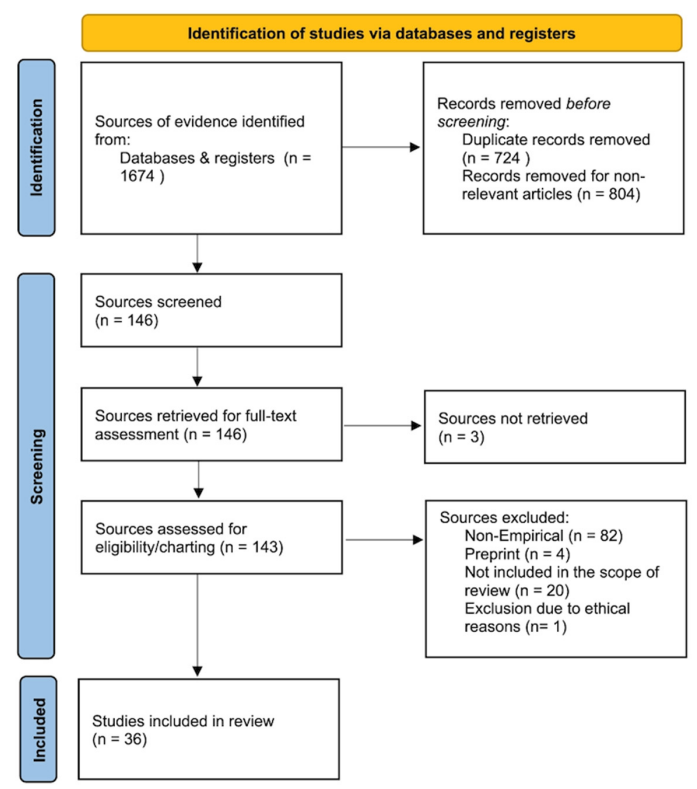}
    \caption{PRISMA-ScR flow chart.}
    \label{fig1}
\end{figure}

\subsection{Search Strategy and Data Extraction}
A customized data-charting form was developed to capture key information: study phase (application scenarios), AI technology, setting, outcomes, and limitations from each included article. It captured key details such as the objectives, AI technologies utilized, main findings, and conclusions. This structured approach allowed for a thorough and organized analysis of the empirical evidence on conversational AI in mental health care. Two reviewers independently extracted all study-level data; any discrepancies were discussed and resolved by consensus adjudication prior to synthesis.

\section{Results}
\subsection{Summary of Reviewed Results}
The review process included 36 articles that showcased survey results of users, clinical students, and mental health professionals’ perceptions toward AI applications. Additionally, all 36 articles evaluated the efficacy of specific aspects of AI applications. To organize the reviewed studies, we classified them based on the primary AI technologies employed. Given the considerable overlap between categories, we focused on highlighting the main types of technologies without reporting precise counts. The technologies identified across the studies include AI chatbots, conversational agents, natural language processing (NLP) tools, large language models (LLMs), machine learning (ML) models, deep learning (DL) models, and AI-based prediction systems. We acknowledge that many studies employed hybrid approaches or combinations of these technologies. To improve clarity, we report general trends and key examples without quantifying the exact number of studies per category. Table 1 lists the technologies mentioned in the articles.

\begin{table}[H]
    \centering
    \caption{AI modalities categories.}
    \begin{tabular}{p{4cm}<\raggedright  p{12cm}<\raggedright }
        \toprule
        Label & Theme \\
        \midrule
        AI Chatbots & Rule-based or scripted dialogue systems that deliver predefined psychoeducation or CBT prompts; no statistical/ML adaptation; always user-interfacing. \\
        Conversational AI Agents & Multi-turn dialogue systems that incorporate traditional NLP components, such as intent classification or sentiment analysis to tailor replies, but do not employ large language model generation. \\
        Machine Learning (ML) Models & Supervised algorithms (e.g., logistic regression, SVM, and gradient boosting) trained on structured or text features to classify diagnosis, predict outcomes, or service flow, generally backend analytics. \\
        Natural Language Processing (NLP) Tools & Standalone language-processing pipelines, tokenization, topic modelling, and emotion detection, used either to support a conversational or analyze text corpora; excludes LLMs. \\
        Large Language Models (LLMs) & Transformer-based generative models with >1 billion parameters (e.g., GPT-3.5/4) capable of free-text generation, contextual memory, and zero-shot reasoning; typically fine-tuned for multi-turn counselling. \\
        Deep Learning (DL) Models & Neural networks such as CNNs or RNNs applied to non-text signals (images, sensor streams) or structured clinical data for pattern recognition and outcome prediction. \\
        AI Prediction Modelling (RPM) & Any ML or DL model embedded in a remote patient-monitoring pipeline that ingests physiological or behavioral signals to stratify risk or trigger alerts in real time. \\
        \bottomrule
    \end{tabular}
    \label{tab1}
\end{table}

This classification directly supports RQ1, helping us map which AI modalities power interventions across different mental health phases. In the following Results sections, we unpack these mappings phase by phase, for example, noting the prominence of rule-based chatbots in screening, NLP agents in empathic support, ML/DL models in post-treatment risk assessment, and emerging LLM agents in multi-turn counseling.
\begin{table}[H]
    \centering
    \caption{Clinical phases of AI deployment in mental health care. Functional roles of AI in mental health interventions.}
    \begin{tabular}{p{4cm}<\raggedright p{12cm}<\raggedright }
        \toprule
        Clinical Phase/Scenario & Descriptions \\
        \midrule
        Pre-treatment/Screening & Interventions used before formal care begins, including online self-referral, triage, or risk screening. \\
        Treatment & AI components integrated during active psychotherapy, pharmacotherapy, or combined treatment phases. \\
        Post-treatment/Monitoring & AI tools used for follow-up care, symptom monitoring, risk assessment, or treatment adjustment after formal treatment. \\
        General support and prevention & Standalone tools aimed at maintaining well-being, reducing stress, or preventing mental health problems in non-clinical or community populations. \\
        Clinical education & AI tools used to train, assess, or upskill mental health professionals, clinical students, or educators. \\
        \midrule
        Functional Category & Descriptions \\
        \midrule
        Assessment & Structured intake or self-report instruments automated by AI to collect clinical or mental health information. \\
        Diagnosis & Tools designed to output diagnostic labels or severity assessments of mental health conditions. \\
        Patient monitoring & Tools providing continuous or periodic tracking of symptoms, behaviors, or physiological markers. \\
        Treatment outcome prediction & Models that forecast treatment responses, dropout risks, or recovery trajectories. \\
        Mental health counseling/therapy & AI-delivered psychotherapeutic interventions, such as cognitive behavioral therapy (CBT), behavioral activation, or problem-solving therapy. \\
        Mental health treatment (Clinical decision support) & AI systems recommending treatment plans, medication, or therapy adjustments, supporting clinician decision making. \\
        Mental health support & Low-intensity support services, such as psychoeducation, emotional assistance, or peer facilitation without formal therapy claims. \\
        \bottomrule
    \end{tabular}
    \label{tab2}
\end{table}

The reviewed studies reported applications across multiple clinical phases (RQ2): pre-treatment (screening and triage), treatment (therapeutic support), post-treatment (monitoring and follow-up), general mental health support and prevention, and clinical education. Functions included assessment, diagnosis, patient monitoring, treatment outcome prediction, mental health counseling or therapy, clinical decision support, and general mental health assistance. Many studies covered multiple functions, which we discuss in detail in the Results subsection “Key Findings in the Applications in Mental Health Care” by linking these functions to observed clinical outcomes and implementation challenges (see Table 2).

\subsection{Analysis and Synthesis of the Results}
The results of the included studies were synthesized using a narrative synthesis approach. Given the expected heterogeneity in study designs and objectives, this method allows for the identification of overarching themes, discussion of patterns and discrepancies, and a nuanced understanding of the current landscape and potential future directions of conversational AI applications in mental health care (see Table 3).

\begin{table}[H]
    \centering
    \caption{Descriptive table of scoping reviews.}
    \resizebox{\linewidth}{!}{
    \begin{tabular}{p{1.5cm}<\raggedright p{3cm}<\raggedright p{3cm}<\raggedright p{8cm}<\raggedright  p{8cm}<\raggedright }
        \toprule
        Reference & Scenario/Application & AI Technology & Purpose & Main Result \\
        \midrule
        \cite{21} & Mental health (MH) assessment & AI chatbot & Examined “Limbic Access” AI in enhancing mental illness recovery in NHS services. & The use of AI tool was associated with an increase in recovery rates from 47.1\% in the pre-implementation period to 48.9\% post-implementation. \\
        \cite{23} & Patient monitoring & AI prediction modelling & AI-based RPM model incorporates RFID for monitoring mental health, targeting vital signs and activity classification. & The implementation can help to monitor patients with mental illnesses sufficiently and effectively support the treatment teams to provide timely interventions, improve patient safety, and prevent incidents such as self-harm. \\
        \cite{31} & Treatment outcomes prediction & Machine learning (ML), RNN & AI applications in iCBT predict mental health outcomes. & The developed AI models demonstrated good accuracy in predicting patient outcomes, resulting in approximately 87\% accuracy after three clinical reviews. \\
        \cite{32} & Detects and diagnoses & ML & AI tool improves MHM assessment, using fewer questions while maintaining diagnostic precision. & The system provided an accuracy level of 89\% in diagnosing mental disorders with only 28 questions asked during diagnostic sessions, reducing questions volume, and encouraging better participation. \\
        \cite{33} & MH counseling/therapy & AI chatbot & AI chatbots’ emotional interactions enhance user satisfaction and retention. & Emotional disclosure in chatbot significantly increases satisfaction and reuse intention. Users’ emotional disclosure intention and the perceived intimacy with a chatbot also mediate the effect. \\
        \cite{34} & Peer support & Conversational AI agent & HAILEY AI system fosters empathy in text-based peer mental health support. & The study showed a substantial growth of users’ empathic response due to the human–AI collaboration. \\
        \cite{23} & MH support & Large language model (LLM) & ChatCounselor, an AI tuned with real counseling data, evaluated for mental health aid. & ChatCounselor demonstrated improved performance in a counseling-specific benchmark, showing promising potential in providing mental health support. \\
        \cite{28} & MH support & LLM, natural language processing (NLP) & Studies chatbots’ personalities, like Big Five Traits, on user engagement in mental health. & The efficacy of chatbots with high conscientiousness drives user engagement, and variations in preferences indicating a match between chatbot and user personalities could influence engagement. \\
        \cite{27} & MH support & Various, mainly NLP & Survey reveals conversational AI’s acceptance for mental health among students. & The survey result highlights the increasing awareness and positive perception among students toward the use of conversational AI technologies for mental health support. \\
        \cite{29} & Diagnosis, MH support & Conversational AI agent & Developed a conversational AI agent that acts like a human therapist, providing accessible emotional support and mental health diagnosis. & The logistic regression model produced improved accuracy of emotion prediction since the chatbot was able to effectively interact with users and quickly respond. \\
        \cite{35} & Diagnosis & Deep learning model & Developed an early diagnostic system for detecting depression by analyzing textual data using deep learning techniques. & The model achieved 99\% accuracy in detecting depression from textual data, outperforming other frequency-based deep learning models. \\
        \cite{36} & MH counseling/therapy, patient monitoring & NLP & AI chatbot with Behavioral Activation enhances mood in users. & The pilot study demonstrated effectiveness in mood improvement, with significant improvements observed in mood scores from pre-usage to post-usage. \\
        \cite{37} & MH support & Conversational AI agent & Study on personality-adaptive conversational agents in mental health care outlines benefits and risks. & The study highlighted both potentials in utilizing PACAs to provide accessible mental health support and serious concerns regarding trust and privacy, indicating requirements. \\
        \cite{38} & Clinical education & ML & Clinical psychology students show strong interest in AI/ML education. & Identifies high interest and perceived importance among students for AI/ML in their education, with high needs for greater formal instruction in these domains. \\
        \cite{39} & Patient referral and clinical assessment & AI chatbot & Conversational AI “Limbic Access” boosts psychotherapy efficiency and patient outcomes. & Improved clinical efficiency and reduced the time that clinicians spent on assessments; improved patient outcomes such as shorter wait times, lower dropout rates, more accurate treatment allocation, and higher recovery rates. \\
        \cite{40} & MH support & AI chatbot, NLP, ML & RCT evaluates “Tess” AI in reducing depression and anxiety in college students. & Showed significant decline in the symptoms of depression and anxiety, which indicated increased engagement and satisfaction with control group. \\
        \cite{25} & MH counseling/therapy & AI chatbot & VR empathy chatbot for college stress shows stress and sensitivity score reductions. & Indicated a reduction in mean stress level and psychological sensitivity scores of the subjects following the intervention by the VRECC system. \\
        \cite{41} & MH counseling/therapy, MH support & AI chatbot, NLP & Focuses on the challenges and needs of moderating digital interventions and counseling offered at Kooth. & Outlines some of the main challenges, including controlling time, interpreting user communication, and addressing hidden risks. 
\end{tabular}}
\end{table}
        \begin{table}[H]
    \centering 
    \resizebox{\linewidth}{!}{
    \begin{tabular}{p{1.5cm}<\raggedright p{3cm}<\raggedright p{3cm}<\raggedright p{8cm}<\raggedright  p{8cm}<\raggedright }
        \cite{42} & MH monitoring & Deep learning model & AI-edge computing system monitors negative info’s impact on pandemic-related mental health. & The system is capable of measuring negative information and its influence on well-being, offering an effective strategy for mental health monitoring during pandemic. \\
        \cite{22} & MH treatment & ML & Assessed the feasibility of recommender systems in personalizing treatment within digital mental health therapy. & Suggested that recommender systems could successfully individualize therapeutic recommendations and improve outcomes. \\
        \cite{43} & Diagnosis, treatment outcomes prediction & Deep learning model & M2C model in art therapy accurately predicts stress levels through art analysis. & The M2C model had 88.5\% accuracy in predicting stress levels based on the data from offered art psychotherapy tests. \\
        \cite{44} & MH counseling/therapy & AI chatbot & AI-driven counselor’s artificial empathy compared with human counselors. & AI-based empathetic counseling was rated with less helpfulness. \\
        \cite{45} & Clinical education & ChatGPT & ChatGPT’s effectiveness in simulating client interactions for counseling practice assessed. & Suggested that while ChatGPT can simulate various aspects of client interactions, it has limitations, such as lack of non-verbal cues. \\
        \cite{46} & Diagnosis, MH support & ChatGPT & ChatGPT’s emotional understanding in conversations evaluated using LEAS. & ChatGPT’s emotional awareness was significantly higher compared to human norms, with improvement over time. It achieved high accuracy in fitting emotions to scenarios. \\
        \cite{26} & MH treatment & ChatGPT & ChatGPT’s response adaptability for BPD and SPD conditions analyzed. & For BPD, ChatGPT showed the ability to respond with a higher emotional resonance than SPD, which suggests its potential use as personalized mental health intervention. \\
        \cite{47} & MH treatment & ChatGPT & Comparison of ChatGPT’s and physicians’ depression treatment recommendations. & ChatGPT recommended psychotherapy more frequently than physicians for mild depression and aligned with guidelines for severe depression treatment, showing no biases. \\
        \cite{48} & MH support & Conversational AI agent & A mental health app design involving participatory protocol with therapists and users for stress management. & Indicated notable symptom improvement and therapist endorsement for AI app integration, highlighting patient engagement benefits. \\
        \cite{49} & Various, mainly MH treatment & Various & Explores mental health professionals’ views on the needs of AI in optimizing care. & Showed themes on practice change, readiness, and educational needs for faster AI adoption, organizational optimizing care with AI. \\
        \cite{50} & MH counseling/therapy & AI chatbot & Examined AI therapist MYLO’s test among youth with mental health issues. & Feedback was positive, noting decreased distress and improved goal conflict resolution. \\
        \cite{51} & MH support & ML & Use machine learning to enhance mental health hotline efficiency by optimizing caller-to-counselor routing. & Significantly increased call volume handled and improved chat quality, outperforming traditional methods, notably during peak times. \\
        \cite{52} & MH counseling/therapy & LLM & Evaluates ChatGPT 3.5’s response to severe depression and suicidal tendencies in simulations. & Agents escalated cases or shut down at critical risk; some provided crisis resources. \\
        \cite{53} & Diagnosis & AI chatbot & Research on DEPRA chatbot’s depression detection effectiveness in Australians. & DEPRA effectively identified depression levels; high satisfaction reported. \\
        \cite{54} & MH support & Conversational AI agent & Assesses Lumen, a voice-based virtual coach for problem solving, through user experience, workload, and interviews. & Users found Lumen highly usable and engaging, despite some feeling rushed in sessions. \\
        \cite{55} & MH counseling/therapy & Conversational AI agent & Study assessed TEO’s effect on aging workers’ stress and anxiety. & No major differences in symptom reduction across groups; however, TEO with therapy improved well-being. \\
        \cite{56} & MH treatment & NLP & Analyzed NLP model biases in psychiatry across demographics, examining health inequality risks. & Underscored AI’s role in enhancing patient management via administrative tasks, real-time support for clinicians, and personalized care through digital phenotyping. \\
        \cite{57} & MH treatment & ML, NLP & Explores AI to enhance patient flow in NHS mental health units, through literature and expert insights, targeting administrative and clinical efficiency. & Underscored AI’s role in enhancing patient management via administrative tasks, real-time support for clinicians, and personalized care through digital phenotyping. \\
        \bottomrule
    \end{tabular}}
    \label{tab3}
\end{table}

\subsection{Key Findings in AI Technologies in Mental Health Care}
This section directly addresses RQ1 by cataloguing the primary AI modalities, rule-based chatbots, traditional NLP agents, ML/DL predictive models, and LLM-based agents, and summarizing the key empirical outcomes and limitations reported for each algorithm class.

\subsubsection{Applying Natural Language Processing}
Through the applications developed by natural language processing (NLP), machine learning (ML), and deep learning (DL), AI tools empower clinical services and treatments and make mental health services much more accessible \cite{25,35}. First, NLP, with mainly applications of chatbots and AI agents, is extensively employed to enhance the interactivity and efficacy of using conversational agents to help deliver mental health care \cite{25,37,40}. Many research-oriented AI-driven chatbots or agents, such as “Hailey”, “MYLO”, and “Limbic Access”, are utilizing sophisticated NLP algorithms to deeply understand human language and initiate, respond, and engage in meaningful conversations related to users’ well-being. Conversations empowered by NLP techniques, such as emotion detection and sentiment analysis, can effectively help provide mental health support and offer computerized therapies \cite{29,36}.

The capabilities of NLP in engaging users through text and voice interactions foster mental health interventions with no limits on time and space. Many AI-based chatbots leveraged NLP to provide mental health support through conversations \cite{23,40}. An AI-enhanced cognitive behavioral therapy (CBT) chatbot named Hailey used NLP to analyze user text inputs and, specifically, detect emotions and train users to give empathic responses for facilitating peer-to-peer communications, and an agent called TEO provided similar techniques to enhance stress reduction \cite{34,48}.

Moreover, preliminary research on ChatGPT highlighted LLM-based conversational AI’s potential to expand access to mental health services to all populations due to its incredible capabilities in understanding human language. It surpasses capabilities in identifying emotions, tailoring interventions for conditions such as borderline personality disorder (BPD) and sensory processing disorder (SPD), and offering advice that aligns with primary health care guidelines for depression. In addition to the opportunities NLP brought, some NLP models in psychiatry still demonstrated significant biases related to religion, race, gender, nationality, sexuality, and age, which highlighted the need for further enhancing the technology \cite{56}.

\subsubsection*{Applying Machine Learning}
Machine learning (ML) has proven to be a powerful tool in predicting mental health outcomes and enhancing diagnostic accuracy, ultimately aiming to improve treatment efficiency and recovery outcomes. By leveraging predictive analytics and classification models, ML provides clinicians with valuable decision-support systems and enables the creation of personalized treatment plans for clients \cite{24,57}. For example, machine learning models, including logistic regression, ridge regression, and LASSO regression, have been used to develop AI-based assessment tools that can accurately predict mental disorders based on responses to a mental health assessment tool like the SCL-90-R \cite{32}. This tool can diagnose mental disorders with an impressive accuracy of 89\% using just 28 questions. Regarding remote patient monitoring, ML algorithms analyze data from devices monitoring vital signs and physical activity to identify subtle changes indicative of worsening mental health conditions, such as depression or anxiety \cite{24,36}. This capability is crucial for timely intervention in these conditions, where even minor changes can be serious indicators of patient distress.

ML has also been applied to optimize operational efficiency in mental health services. For instance, it has been used to improve caller-counselor matching in a mental health contact center, leading to more efficient use of resources and higher quality of service \cite{51}. Furthermore, in larger health care organizations, ML algorithms have been used to streamline patient flow, which indirectly contributes to the improvement of mental health care \cite{57}. In therapeutic settings, ML offers valuable insights into the efficacy of interventions on an individual basis. By analyzing transcripts of therapy sessions, patient feedback, and progress over time, ML models can assist therapists in personalizing their approaches to meet the needs of each patient \cite{22,31}. This ability to tailor treatment plans based on data-driven insights has the potential to enhance the effectiveness of mental health interventions significantly.

\subsection{Applying Deep Learning}
Deep learning (DL) algorithms, a subset of machine learning, learn complex patterns directly from data, enabling accurate predictions and analyses for innovative applications such as real-time emotional state monitoring and predictive analytics for treatment outcomes. DL contributes to improving online cognitive behavioral therapy (CBT) and art psychotherapy by customizing mental health treatments to cater to individual needs \cite{31,43}. In the realm of iCBT, DL algorithms and recurrent neural networks (RNNs) are employed to analyze anonymous patient data \cite{31}. They detect patterns that accurately forecast treatment outcomes, assisting in identifying mental health issues and the customization of therapy for more effective and personalized interventions \cite{31,35}.

Similarly, in art psychotherapy, DL models with co-attention mechanisms are revolutionizing the evaluation of art therapy \cite{43}. These models assess stress and mood levels based on multiple data points. DL’s capacity to interpret complex emotional expressions and provide insights that align closely with therapeutic goals \cite{42,43}. Through these advancements, DL serves as a technological tool and acts as a bridge to compassionate, precise, and personalized mental health care. Having mapped the core AI modalities above, we now turn to how these tools (chatbots, predictive models, and LLMs) are deployed at each phase of mental health care.

\subsection*{Key Findings in the Applications in Mental Health Care}
Here, we answer RQ2 by showing how those same AI modalities are deployed across the five clinical phases (pre-treatment, treatment, post-treatment monitoring, clinical education, and prevention), highlighting which technologies are most prevalent in each phase and where critical gaps remain.

The review identified various applications of AI in mental health care that cater to diverse users, ranging from patients and clinicians to the general public and psychology students. These applications serve different purposes, including providing computerized therapies to patients, offering early-stage mental health support, assisting clinicians with diagnosis and treatment, and enhancing learning for psychology students. Overall, the studies showcased the positive impact of AI technology in improving mental health care and emphasized the significant potential of these applications to revolutionize the industry. Interestingly, most of the studies focused on how these AI-powered applications can complement and enhance the existing services provided by clinicians rather than replacing them. Figure 2 illustrates the four pillars of AI applications in mental health care, demonstrating their extensive utilization across various stages of support and treatment. The four pillars encompass four key stages: pre-treatment, treatment, post-treatment, and general improvement and prevention. Artificial intelligence is leveraged throughout these stages to enhance various aspects of mental health services, creating a continuous source for optimized care and support. In the pre-treatment phase, AI expedites assessment, facilitates initial diagnosis, and aids in referral. During treatment, AI refines diagnoses, personalizes treatment plans, predicts outcomes, and delivers AI-based therapeutic interventions. Post-treatment involves leveraging AI for remote monitoring and risk evaluation. The general improvement and prevention stage focuses on providing proactive mental health support to the broader population through AI-enabled resources.

\begin{figure}[H]
    \centering
    \includegraphics[width=0.8\textwidth]{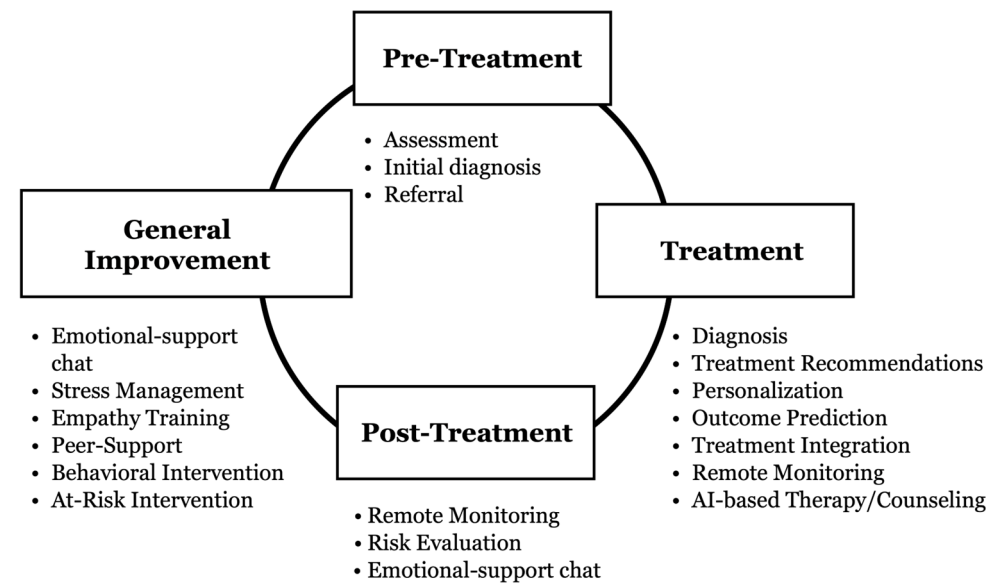}
    \caption{Four pillars framework where AI-driven interventions are deployed.}
    \label{fig2}
\end{figure}

Throughout the four pillars framework, AI technologies benefit multiple stakeholders (see Figure 3). Patients gain access to fast-track services and more effective, personalized interventions. Clinicians can make enhanced, data-driven decisions. Health organizations improve efficiency and treatment efficacy. The general public benefits from increased access to low-cost, AI-powered mental health resources. This integrated model showcases the vast potential of AI in revolutionizing mental health care. In the following sections, we turn to the existing literature to explore examples of AI implementation within each stage of the model.

\begin{figure}[H]
    \centering
    \includegraphics[width=0.8\textwidth]{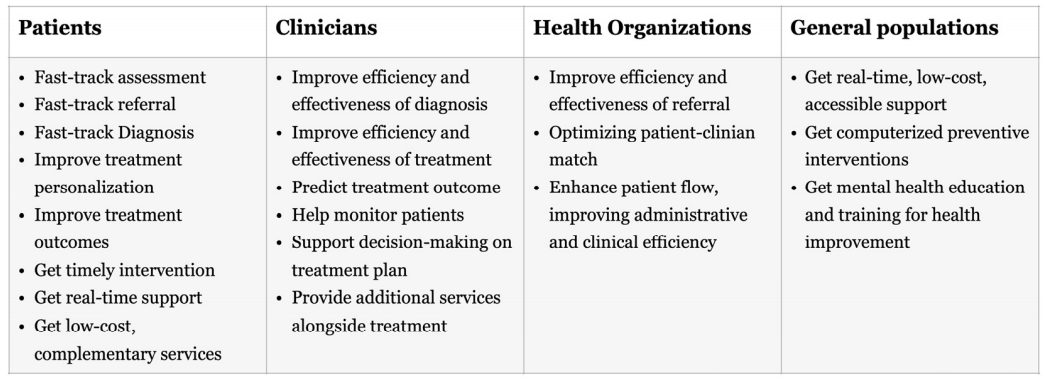}
    \caption{Benefits of AI technologies for multiple stakeholders in health care.}
    \label{fig3}
\end{figure}

\subsection{Applications in the Pre-Treatment Stage}
First, the AI chatbot “Limbic Access” has shown great promise in the pre-treatment stage of mental health care. This tool, developed by Shaik et al. \cite{24} and further refined by Rollwage et al. \cite{39}, helps clinicians assess patients and refer them to appropriate services.

One key feature of Limbic Access is its AI self-referral tool. This user-friendly tool collects important information from patients applying for the UK’s National Health Service (NHS) Talking Therapy program. It asks about the patient’s symptoms using standardized questionnaires like the Patient Health Questionnaire-9 (PHQ-9) and the Generalized Anxiety Disorder Assessment-7 (GAD-7). The tool also gathers demographic details and other clinical information. By automating this intake process, Limbic Access makes it faster and easier for clinicians to assess patients and get them started on the right track.

Rollwage et al. evaluated the impact of Limbic Access by comparing outcomes from patients who used the AI tool with those who followed traditional assessment methods \cite{39}. The results were promising. Patients who used Limbic Access waited less time for both clinical assessment and treatment. They were also less likely to drop out of the program and more likely to recover overall. As a result, Limbic Access has proven to be a valuable asset in the pre-treatment stage, streamlining the assessment process and improving patient outcomes.

\subsection{Applications in the Treatment and the Post-Treatment Stage}
The effectiveness of AI in assisting the treatment process is highlighted by its ability to aid in the detection and diagnosis of conditions, predict treatment outcomes, and monitor patient progress \cite{32,36,41,43,53}. AI systems have achieved high accuracy in diagnosing mental health disorders and predicting stress levels from art therapy tests \cite{32,43,53}.

AI also enhances treatment outcome predictions, leading to better personalization and planning \cite{24,31}. By analyzing non-invasive digital data, identifying patterns in mental health symptoms, and considering past preferences and ratings, AI provides strong support for clinicians to deliver timely and personalized interventions \cite{22,24,31}. The accuracy of AI in predicting outcomes and the continuous improvement of algorithms demonstrate its effectiveness in personalizing treatment and improving outcomes \cite{22,26,35,47}. In terms of clinical knowledge, ChatGPT exhibits high capabilities in providing treatment recommendations and aligning well with clinical guidelines \cite{47}. AI-enabled Remote Patient Monitoring (RPM) systems support clinicians in delivering timely interventions, enhancing patient safety, and preventing incidents like self-harm \cite{24}.

AI agents like Tess and MYLO have proven they can deliver therapy and counseling services that rival human clinicians, showing moderate to high efficacy in reducing symptoms, improving health, and increasing user satisfaction \cite{25,31,33,36,40,48,50,55}. These agents are trained to use emotions and empathy to provide high-quality services, and users report increased satisfaction when these agents disclose emotions \cite{33,44,50,55}. However, users still prefer human responses over AI empathy \cite{33,44}.

Research has explored how AI agents’ emotions, empathy, and personalities impact intervention outcomes and satisfaction. While some aspects have positive effects, like driving user engagement, there is room for improvement in advancing models and algorithms to better meet the high demand for counseling \cite{33,44}. AI agents demonstrate flexibility in format, offering services through text, chat, videos, and even virtual reality \cite{25,36,51}. They can also support specific mental health therapies like CBT, BA, problem-solving therapy, and problem-solving treatment \cite{31,36,48,54}. Implementing BA in a chatbot has shown improvements in mood levels \cite{36}.

Since therapies can be standardized, AI offers opportunities to automate services and make them widely accessible through digital applications. However, challenges exist, such as risk assessment and ensuring timely human intervention for individuals in crisis \cite{22}. Additionally, some AI systems, like ChatGPT, discontinue interventions for high-risk conditions and cannot provide effective referrals \cite{52}. This highlights the need for human monitoring of AI-based interventions to identify high risks and establish systems for timely referrals and crisis interventions \cite{22}.

\subsection{Applications in General Support, Improvement, and Prevention}
Beyond the treatment period, AI provides ongoing support to clients and offers general mental health resources for improvement and prevention \cite{23,37}. Numerous AI chatbots and platforms are designed to provide accessible, cost-effective, and scalable mental health support outside of the clinical setting \cite{23,34,36,37,40,41,44,50}.

Chatbots can identify human emotions and sentiments and assess psychological conditions to generate individualized support, with studies showing high accuracy in measuring users’ psychological states \cite{29,35}. AI agents have demonstrated significantly higher emotional awareness than human norms and can provide useful interventions to low- and middle-risk conditions, ensuring safety \cite{46,52}. For example, Tess, a mental health support chatbot, reported significant reductions in symptoms of depression and anxiety among university students, demonstrating higher engagement and satisfaction compared to a control group \cite{40}. Another chatbot, HAILEY, focused on improving users’ empathetic abilities and found substantial results in increasing users’ empathetic responses \cite{25}.

Research is ongoing to enhance AI's models, algorithms, and product design to deliver mental health services better \cite{23,29}. For instance, ChatCounselor developed a fine-tuned model by analyzing real counseling data and established quality benchmarks, outperforming all open-source models \cite{23}. These advancements present exciting opportunities to support the mental health of many populations, particularly those without access to traditional clinical care. For example, an Israeli mental health hotline applied AI to provide daily support to various populations \cite{51}.

AI-driven tools also play a crucial role in stress management and emotional support for the general public. A voice-based virtual coach called Lumen offers problem-solving support through interactive sessions, with users reporting high usability and engagement \cite{54}. Similarly, personality-adaptive conversational agents (PACAs) are being developed to provide tailored mental health support, though concerns about trust and privacy remain important considerations \cite{37}. These tools demonstrate the potential of AI to make mental health support more accessible and personalized for diverse users.

\subsection{Applications in the Treatment and the Post-Treatment Stage}

The effectiveness of AI in assisting the treatment process is highlighted by its ability to aid in the detection and diagnosis of conditions, predict treatment outcomes, and monitor patient progress \cite{32,36,41,43,53}. AI systems have achieved high accuracy in diagnosing mental health disorders and predicting stress levels from art therapy tests \cite{32,43,53}.

AI also enhances treatment outcome predictions, leading to better personalization and planning \cite{24,31}. By analyzing non-invasive digital data, identifying patterns in mental health symptoms, and considering past preferences and ratings, AI provides strong support for clinicians to deliver timely and personalized interventions \cite{22,24,31}. The accuracy of AI in predicting outcomes and the continuous improvement of algorithms demonstrate its effectiveness in personalizing treatment and improving outcomes \cite{22,26,35,47}. In terms of clinical knowledge, ChatGPT exhibits high capabilities in providing treatment recommendations and aligning well with clinical guidelines \cite{47}. AI-enabled Remote Patient Monitoring (RPM) systems support clinicians in delivering timely interventions, enhancing patient safety, and preventing incidents like self-harm \cite{24}.

AI agents like Tess and MYLO have proven they can deliver therapy and counseling services that rival human clinicians, showing moderate to high efficacy in reducing symptoms, improving health, and increasing user satisfaction \cite{25,31,33,36,40,48,50,55}. These agents are trained to use emotions and empathy to provide high-quality services, and users report increased satisfaction when these agents disclose emotions \cite{33,44,50,55}. However, users still prefer human responses over AI empathy \cite{33,44}.

Research has explored how AI agents’ emotions, empathy, and personalities impact intervention outcomes and satisfaction. While some aspects have positive effects, like driving user engagement, there is room for improvement in advancing models and algorithms to better meet the high demand for counseling \cite{33,44}. AI agents demonstrate flexibility in format, offering services through text, chat, video, and even virtual reality \cite{25,36,51}. They can also support specific mental health therapies like CBT, BA, problem-solving therapy, and problem-solving treatment \cite{31,36,48,54}. Implementing BA in a chatbot has shown improvements in mood levels \cite{36}.

Since therapies can be standardized, AI offers opportunities to automate services and make them widely accessible through digital applications. However, challenges exist, such as risk assessment and ensuring timely human intervention for individuals in crisis \cite{22}. Additionally, some AI systems, like ChatGPT, discontinue interventions for high-risk conditions and cannot provide effective referrals \cite{52}. This highlights the need for human monitoring of AI-based interventions to identify high risks and establish systems for timely referrals and crisis interventions \cite{22}.

\subsection{Applications in General Support, Improvement, and Prevention}

Beyond the treatment period, AI provides ongoing support to clients and offers general mental health resources for improvement and prevention \cite{23,37}. Numerous AI chatbots and platforms are designed to provide accessible, cost-effective, and scalable mental health support outside of the clinical setting \cite{23,34,36,37,40,41,44,50}.

Chatbots can identify human emotions and sentiments and assess psychological conditions to generate individualized support, with studies showing high accuracy in measuring users’ psychological states \cite{29,35}. AI agents have demonstrated significantly higher emotional awareness than human norms and can provide useful interventions to low-and middle-risk conditions, ensuring safety \cite{46,52}. For example, Tess, a mental health support chatbot, reported significant reductions in symptoms of depression and anxiety among university students, demonstrating higher engagement and satisfaction compared to a control group \cite{40}. Another chatbot, HAILEY, focused on improving users’ empathetic abilities and found substantial results in increasing users’ empathic responses \cite{25}.

Research is ongoing to enhance AI’s models, algorithms, and product design to deliver mental health services better \cite{23,29}. For instance, ChatCounselor developed a fine-tuned model by analyzing real counseling data and established quality benchmarks, outperforming all open-source models \cite{23}. These advancements present exciting opportunities to support the mental health of many populations, particularly those without access to traditional clinical care. For example, an Israeli mental health hotline applied AI to provide daily support to various populations \cite{51}.

\subsection{Clinical Education}

AI has proven its potential to support mental health care by aiding in the education of professionals. A study surveyed clinical psychology students from a Swiss university and found that students recognized the growing importance of AI in mental health care and expressed a significant interest in and need for instruction on AI in their clinical training \cite{38}.

A descriptive study of mental health professionals emphasized the need to adopt AI in practice and highlighted the importance of educational initiatives for broader adoption \cite{40}. Another study explored the feasibility of using ChatGPT to simulate client scenarios and help clinical students practice their counseling skills \cite{45}. While ChatGPT has limitations, such as a lack of non-verbal cues or overly idealized situations, it demonstrated capabilities in displaying authenticity, consistency, appropriate emotional expression, cultural sensitivity, and empathy in simulations, presenting its effectiveness for clinical training \cite{45}.

ChatGPT’s emotional awareness for aligning with psychotic symptoms indicated its feasibility as a useful research tool for understanding emotional experiences in psychopathology \cite{26}. Furthermore, ChatGPT’s proven ability to provide unbiased treatment recommendations with clinical guidelines can effectively support clinical students in their educational journey \cite{46}.

\section{Discussions}

Unlike recent systematic reviews focused exclusively on generative AI or ethical risks, our mapping spans rule-based chatbots through LLMs and charts empirical outcomes across five clinical phases. This review demonstrates AI applications with different efficacy and potential, simply displaying the different levels of effectiveness and applicability.

\subsection{AI Modalities Applied Across Phases}

First, AI is widely used as an emotional support agent and clinical assistant, most often delivered through chatbots \cite{14,25,36}. Many cases of baseline emotional support and clinical assistance reported high efficacy and involved the lowest level of risks; thus, these applications can be applied most quickly and universally \cite{58}. It is foreseeable that an emotional support chatbot will be a core AI application for researchers paying attention in the short term, and AI can start its application by assisting clinical work before the treatment stage. Secondly, the main context in AI, which might have a relatively high efficacy and potential in mental health care, mainly focuses on supporting the clinical work in the real clinical setting. The strengths of AI in supporting treatment displayed high potential in improving treatment outcomes by developing personalized treatment, providing treatment recommendations and diagnoses, and predicting treatment outcomes \cite{21,22,29}. Thirdly, AI can complete partial therapies or counselling sessions, automating the treatments. Using AI to offer therapies or counselling sessions directly can potentially solve the problem of accessibility of mental health care \cite{43,59,60}. Meanwhile, the significant risks and ethical concerns make this layer hard to realize in the short term \cite{34,40,41,42,52}. It is expected to advance models and algorithms to improve accuracy and reduce biases. Pilot tests and clinical trials should be done along with the development and iteration processes.

\subsection{AI-Clinician Collaboration}

All applications are perceived as supporters of human clinicians rather than replacements, at present with their proven track record for effectively supporting clinicians’ work \cite{32,42,43,47,57}. A pivotal question of deploying AI in mental health care revolves around the work division and integration of AI and human clinicians, which would largely influence the future direction. This panel raises three main questions for debate. The first concerns the respective responsibilities of AI and human clinicians. The second addresses how human clinicians integrate AI into their existing workflow and the actual impact. The third is about the ethical and regulatory obligations of humans to oversee AI’s activities. Regarding the first question, AI is currently viewed as a complement to human clinicians’ existing jobs, but it will have a higher potential to replace parts of clinicians’ jobs, including diagnosis and providing therapies \cite{24,29}. At the current stage, there is a lack of views seeing AI as a standalone method because of its high risks in resolving accuracy; instead, AI is boosting clinicians’ efficiency as a tool \cite{21,24,27,31,32,34,36}.

The second question focuses on the integration of AI into clinical services. This will depend on organizational adaptability and will face concerns from various stakeholders, such as clinicians, management, board members, shareholders, and patients \cite{61,62}. For example, if human clinicians are highly resistant to adopting specific AI applications, integration cannot go smoothly. The management and board might reject the integration to establish the AI transformation as highly risky and likely to cause public concerns. To see organizational readiness for AI adaptation, it is crucial to adopt actions like AI literacy and application training, conduct pilot tests to demonstrate benefits, and hold organizational meetings for discussion of concerns. At this stage, there is no universally recognized way to integrate AI into clinical settings, and it will largely depend on how organizations practically lead changes. Detailed standards will be developed as industrial organizations, academic associations, and policy authorities provide compliance guidelines.

The third question addresses the responsibility of overseeing AI’s risks and problems. Currently, AI cannot be solely held accountable, so human clinicians and organizations will be responsible for overseeing any risks and problems caused by AI. This will increase the workload of clinicians and require a higher level of technological literacy. When organizations apply AI applications in the clinical setting, they should provide comprehensive training to clinicians, and develop a scheme for monitoring and accountability management \cite{34,40,41,42,52}. As AI becomes more accurate and reliable, the responsibility of overseeing will decrease. However, in the short term, clear definitions of accountability will be needed.

\subsection{Strengths, Weaknesses, Opportunities, and Threats}

This section directly addresses RQ3 by synthesizing the key strengths, weaknesses, opportunities, and threats of AI-driven digital interventions as reported across our 36 included studies. AI technologies have several key strengths that hold promise for mental health care. First, they can increase access to services by overcoming economic, geographical, and logistical barriers. Chatbots like Limbic Access can provide first-line support to anyone with an internet connection, democratizing access to mental health resources \cite{21}. AI-powered chatbots also offer personalized experiences, elevating user satisfaction in counseling \cite{33}. Some conversational agents, like HAILEY, are designed to foster empathy in peer support, potentially creating meaningful connections for users \cite{34}. From an analytical standpoint, machine learning (ML) can uncover insights from large mental health datasets, and ML models are being used to predict treatment outcomes in internet-based therapies \cite{31}. Finally, large language models like ChatCounselor, fine-tuned with actual counseling data, present a scalable path for providing support to a large number of users simultaneously \cite{23}.

Despite these strengths, AI technologies also have several weaknesses. One of the most pressing concerns is privacy and security. AI chatbots hold sensitive patient information, and their implementation poses data security challenges \cite{39}. There is also a risk of misinterpretation by AI models, highlighting the need for human oversight to prevent errors \cite{63}. Questions remain about whether AI tools truly enhance assessment efficiency without compromising diagnostic precision \cite{64}. For instance, can AI agents like HAILEY genuinely foster empathy, and can chatbots with certain personalities effectively engage users? These questions are still under debate \cite{28,34}. Technically, AI is still far from effectively recognizing mental disorders, posing a significant limitation to its implementation \cite{63}.

The field of AI in mental health is rapidly evolving, presenting several opportunities for growth and improvement. AI prediction modelling could refine patient monitoring, enabling earlier interventions and better health outcomes \cite{63}. There is optimism about large language models creating novel, large-scale mental health solutions \cite{65}. For instance, researchers are developing interpretable mental health analysis tasks using generative models like LLMs \cite{63}. Collaboration between stakeholders, including patients and AI researchers, is crucial for human-centered mental health AI research, presenting an opportunity to develop tools that meet real user needs \cite{65}.

Despite these opportunities, several threats must be addressed to realize the potential of AI in mental health. Algorithmic bias could exacerbate existing health disparities, undermining the equity of AI-powered mental health tools \cite{66}. The need to regulate and monitor AI-based mental health apps is pressing, and navigating these challenges while harnessing AI’s strengths is crucial for its successful integration into care. From a technical standpoint, the need for ongoing model enhancement to prevent accuracy decline is clear \cite{66}. The field of ML is rapidly evolving and requires ongoing assessment to ensure tools remain effective and safe \cite{63,66}. Failure to address these threats could hinder the adoption of AI technologies in mental health care. Recent models do not systematically disadvantage minority groups.

\subsection{Advancing Mental Health Prevention and Improvement}

In responding to RQ4, we outline which AI modalities and applications show the greatest maturity for population-level prevention and ongoing support, and we set the stage for the following sections, Policy Implications, which further elaborate the strategic priorities and barriers identified.

It is expected that many AI applications have a high potential to advance mental health prevention and improvement. Prevention is one of the biggest opportunities AI brings because AI can generate interventional services for a huge number of populations through emotional support chatbots, behavioral intervention apps, and peer support platforms \cite{23,36,65}. All these services can effectively improve mood levels and reduce stress levels \cite{23,34,36,37,40,41,44,50}. AI’s capacity to analyze large datasets can uncover patterns and risk factors that were previously hidden, allowing for more individualized and efficient early intervention strategies \cite{24,23} In addition, the predictive analytics of AI can improve screening processes, making them more effective and less invasive, thus increasing the number of people who take part in preventive measures \cite{39,57}. The combination of AI and mobile health apps with wearable devices provides opportunities for continuous monitoring and real-time personalized feedback, thus improving preventive care \cite{67}. Utilizing these low-cost, widely applied AI applications, mental health prevention can be conducted at a population level and designed to adapt to a specific population’s needs. If AI applications can reduce the number of patients with high-level symptoms, the demand for traditional mental health facilities can be reduced, contributing to the global mental health care industry.

It is foreseeable that government agencies can apply AI applications in different populations for prevention and improvement work to reduce the number of at-risk populations, such as students, the elderly, and people with substance use \cite{68,69}. Organizations like schools and local communities can develop and implement specific interventional programs that AI empowers them to meet the specific challenges of each demographic. For example, AI can provide personalized learning and coping mechanisms for students under academic and social pressures, support programs designed for the elderly to face loneliness and cognitive decline, and relapse prevention tools for people with substance use. The integration of AI tools into the current health, educational, and social welfare systems enables these programs to provide adaptive support and interventions in real time \cite{70}. This approach is not only directed at immediate risk reduction but also at the development of mental resilience in the long run, which will result in a reduction of general mental health problems and a move toward a more preventive mental health care paradigm. The following policy section builds on these findings to outline how regulatory frameworks can support the safe and equitable implementation of AI innovations.

\subsection{Policy Implications}

A contemporary policy architecture for AI in mental health care must protect service users while still allowing technology to evolve at pace. This balance can be struck by weaving together three strands (e.g., patient safety, innovation incentives, and bias mitigation) into a single, adaptive regulatory fabric. First, regulators must insist on secure data handling, algorithmic transparency, routine clinical validation, and clear lines of human oversight so that new tools do not erode public trust or amplify harm. Second, they should create pro-innovation pathways, regulatory sandboxes, adaptive licensing, and rapid ethics review that give researchers and companies room to experiment without waiting for blanket legislation to catch up. Third, every approval process must include explicit checks for demographic performance and inclusive training data so that diagnostic road maps show what this regulation-for-innovation ecosystem looks like in practice. New Zealand’s Digital Mental Health and Addiction Service Delivery Framework (DMHAS) advocates a co-regulatory model in which authorities define outcome standards, while innovators choose the technical means to meet them, employing instruments that range from voluntary codes and accreditation seals to binding legislation, each calibrated to the level of risk and the speed of technological change \cite{71}. In a similar vein, the Mental Health Commission of Canada urges policymakers to evaluate AI systems at three translational checkpoints: whether an algorithm that performs well in silo retains its accuracy with real patients, whether it integrates into clinical workflows without adding cognitive burden for clinicians, and whether it scales equitably across regions and demographic groups \cite{72}. Together, these examples suggest a constructive path forward. Policies should facilitate the rapid yet evidence-based migration of laboratory prototypes to bedside tools, provide clinicians with clear operational guidance so that AI augments rather than complicates practice, and embed continuous outcome monitoring so that accountability keeps pace with innovation. A progressive, layered framework, neither technology-neutral nor technology-prescriptive, can therefore foster creativity, safeguard patients, and ensure that AI becomes a feasible, trustworthy ally in mental health services.

\subsection{Limitations}

This review is subject to several constraints that should interpret its findings with caution. First, AI for mental health is a fast-moving field; studies released after our January 2024 cutoff, including work on the latest large language model generations, are not reflected here. Additionally, the search strategy may not have captured all relevant literature. We limited our search to English-language publications, so relevant studies in other databases or non-English sources may have been overlooked.

Second, the heterogeneity of study designs, intervention types, outcome measures, and reporting practices precluded a formal risk-of-bias appraisal and any quantitative synthesis. We opted for a narrative, chart-based approach to preserve breadth, but this means we cannot provide pooled effect sizes or graded levels of evidence. Although the first author screened and extracted data, the assignment of each paper to one of the clinical phases required subjective judgment, introducing a possibility of selection or interpretation biases.

Third, many included studies examined prototypes or pilot deployments in controlled or single-site settings, often with convenience samples. Real-world implementation factors, such as integration into existing clinical workflows, sustainability, and user diversity, were rarely reported in depth. Consequently, the generalizability of reported benefits to routine practice remains uncertain, and future research should prioritize multi-site evaluations, transparent reporting of model development, and rigorous assessments of ethical, privacy, and equity impacts.

Finally, as a scoping review, this study aimed to map the landscape rather than critically appraise study quality. Some included studies may have been preliminary investigations, used correlational designs, or relied on small sample sizes, and no formal quality assessment or risk-of-bias evaluation was conducted. Therefore, findings should be interpreted with caution, and future systematic reviews should incorporate structured quality appraisal to evaluate the robustness and reliability of the evidence base.

\section{Conclusions}

This scoping review synthesized findings from 36 studies on artificial intelligence-driven digital interventions across screening, treatment, monitoring, clinical education, and prevention in mental health care. Our mapping of chatbots, natural language processing, machine learning, deep learning, and large language models highlighted growing evidence of artificial intelligence’s contributions to expanding access, enhancing symptom monitoring, and supporting personalized interventions. At the same time, challenges such as algorithmic bias, data privacy risks, and integration barriers underscore the need for ethical design, transparent model development, and human oversight.

While previous review studies have provided valuable insights into specific aspects of AI in mental health, such as diagnostic applications of ChatGPT, ethical and regulatory discussions, or evaluations of chatbots for anxiety treatment \cite{73,74,75,76,77,80}, these have often focused on single technologies, clinical phases, or conditions. In contrast, the current scoping review offers a broader synthesis by mapping multiple AI modalities across five clinical phases. By linking conceptual insights with empirical outcomes, this review complements and extends prior work, providing an integrated reference for research, practice, and policy.

Future research should prioritize multi-site evaluations, longitudinal studies in diverse populations, and rigorous assessments of safety, privacy, and equity impacts. Collaboration among clinicians, artificial intelligence developers, policymakers, and patients will be essential to ensure artificial intelligence systems are clinically effective, ethically sound, and socially equitable. By offering an empirical map of artificial intelligence applications across the mental health care continuum, this review provides a foundation for guiding research, practice, and policy toward responsible integration of artificial intelligence in mental health services.

\section*{Author Contributions}
Conceptualization, Y.N. and F.J.; methodology, Y.N. and F.J.; formal analysis, Y.N. and F.J.; investigation, Y.N. and F.J.; resources, Y.N. and F.J.; data curation, Y.N. and F.J.; writing—original draft preparation, Y.N. and F.J.; writing—review and editing, Y.N. and F.J.; visualization, Y.N. and F.J.; supervision, F.J.; project administration, F.J.; funding acquisition, F.J. All authors have read and agreed to the published version of the manuscript.

\section*{Funding}
This research received no external funding.

\section*{Institutional Review Board Statement}
Not applicable.

\section*{Informed Consent Statement}
Not applicable.

\section*{Data Availability Statement}
Not applicable.

\section*{Acknowledgments}
The authors would like to thank the academic editors and reviewers for their valuable comments and suggestions.

\section*{Conflicts of Interest}
The authors declare no conflict of interest.


\nocite{*}
\bibliographystyle{unsrt}  
\bibliography{references}

\end{document}